\documentclass[conference]{IEEEtran}
\IEEEoverridecommandlockouts
\usepackage{cite}
\usepackage{amsmath,amssymb,amsfonts,bbm,bm}
\usepackage{algorithm,algpseudocode}
\usepackage{graphicx}
\usepackage{float}
\usepackage{subfigure}
\usepackage{textcomp}
\usepackage{xcolor}
\usepackage{float}
\usepackage{paralist}
\usepackage{multirow}
\usepackage{booktabs}
\usepackage[font=small]{caption}
\usepackage{subcaption}
\usepackage{graphicx}
\usepackage{tikz}
\usetikzlibrary{shapes.geometric, arrows, positioning}
\usepackage[nolist]{acronym}
\usepackage{epstopdf}

\DeclareMathOperator*{\argmax}{argmax}

\newtheorem{remark}{Remark}

\newtheorem{definition}{Definition}

\newcommand{\mP}{\mathbf{P}}

\newcommand{\bu}{\mathbf{u}}

\newcommand{\bx}{\mathbf{x}}

\newcommand{\bp}{\mathbf{p}}

\newcommand{\bdelta}{\boldsymbol{\delta}}

\newcommand{\setR}{\mathbb{R}}

\newcommand{\setL}{\mathbb{L}}

\newcommand{\R}{\mathcal{R}}

\newcommand{\se}{\mathrm{SE}}

\newcommand{\trp}{\mathsf{T}}

\newcommand{\set}[1]{\left\lbrace #1 \right\rbrace}

\newcommand{\brc}[1]{\left( #1 \right)}
\newcommand{\dbc}[1]{\left[ #1 \right]}

\newcommand{\abs}[1]{\left\vert #1 \right\vert}

\newcommand{\Ex}[2]{\mathbb{E}_{#2} \left\lbrace #1 \right\rbrace}
\newcommand{\subto}{\text{ \normalfont s.t.} \;\;}

\def\BibTeX{{\rm B\kern-.05em{\sc i\kern-.025em b}\kern-.08em
    T\kern-.1667em\lower.7ex\hbox{E}\kern-.125emX}}

\begin{document}
\begin{acronym}
\acro{ap}[AP]{access point}
\acro{pass}[PASS]{pinching-antenna system}
\acro{mimo}[MIMO]{multiple-input multiple-output}
\acro{ota-fl}[OTA-FL]{over-the-air federated learning}
\acro{los}[LoS]{line-of-sight}
\acro{ps}[PS]{parameter server}
\acro{noma}[NOMA]{non-orthogonal multiple access}
\acro{irs}[IRS]{reconfigurable intelligent surface}
\acro{snr}[SNR]{signal to noise ratio}
\acro{sinr}[SINR]{signal to interference and noise ratio}
\acro{bcd}[BCD]{block coordinate descent}
\acro{rzf}[RZF]{regularized zero-forcing}
\acro{tdma}[TDMA]{time-division multiple access}
\acro{fp}[FP]{fractional programming}
\acro{mrt}[MRT]{maximal-ratio transmission}
\acro{zf}[ZF]{zero-forcing}
\acro{mse}[MSE]{mean square error}
\acro{ee}[EE]{energy efficiency}
\acro{em}[EM]{electromagnetic}
\end{acronym}

\title{Dynamic and Static Energy Efficient Design of Pinching Antenna Systems}
\author{
\IEEEauthorblockN{Saba Asaad}
\IEEEauthorblockA{University of Toronto\\
saba.asaad@utoronto.ca}
\and
\IEEEauthorblockN{Chongjun Ouyang}
\IEEEauthorblockA{Queen Mary University\\
c.ouyang@qmul.ac.uk}
\and
\IEEEauthorblockN{Ali Bereyhi}
\IEEEauthorblockA{University of Toronto\\
ali.bereyhi@utoronto.ca}
\and
\IEEEauthorblockN{Zhiguo Ding}
\IEEEauthorblockA{Nanyang Technological University\\
	zhiguo.ding@ntu.edu.sg}
\thanks{This work was supported in part by the German Research Foundation (Deutsche Forschungsgemeinschaft - DFG).}
}

\maketitle
\begin{abstract}
We study the energy efficiency of pinching-antenna systems (PASSs) by developing a consistent formulation for power distribution in these systems. The per-antenna power distribution in PASSs is not controlled explicitly by a power allocation policy, but rather implicitly through tuning of pinching couplings and locations. Both these factors are tunable: ($i$) pinching locations are tuned using movable elements, and ($ii$) couplings can be tuned by varying the effective coupling length of the pinching elements. While the former is feasible to be addressed dynamically in settings with low user mobility, the latter cannot be addressed at a high rate. We thus develop a class of hybrid dynamic-static algorithms, which maximize the energy efficiency by updating the system parameters at different rates. Our experimental results depict that dynamic tuning of pinching locations can significantly boost energy efficiency of PASSs.
\end{abstract}
%
%
\section{Introduction}
\label{sec:intro}
Following the DOCOMO demonstration in \cite{suzuki2022pinching}, \acp{pass} have been proposed as a new technology to realize an analog front-end with flexible-antenna arrays \cite{ding2024flexible}. 
A \ac{pass} consists of a long dielectric waveguide in which \ac{em} waves propagate with minimal attenuation. The \ac{em} signal is radiated from this waveguide using the so-called \textit{pinching elements} which are freely moving across the waveguide \cite{LiuPASS}. The long scale of the waveguide allows the elements to place close to users and thereby drastically reduce the path-loss \cite{LiuPASS,ouyang2025array}. Early studies on \ac{pass} have shown that by optimally placing and activating pinching elements, one can push the system toward a \textit{near-wired} behavior, where the path-loss is dominated by the the free-space propagation at the \textit{last-meter} transmission \cite{ding2024flexible}. Motivated by its high degrees of freedom, several recent lines of work have investigated design aspects of \ac{pass}. The work in \cite{bereyhi2025icc} and \cite{bereyhi2025mimopass} studies the integrability of this technology to \ac{mimo} systems, developing low-complexity algorithms for hybrid uplink and downlink beamforming with \ac{pass}. The study in \cite{PassOut} characterizes sevral performance metrics under realistic waveguide attenuation and path-loss models. Enhancing physical layer security using the high flexibility of \acp{pass} is studied in \cite{shan2025securemulti}. 

\subsection{Energy Efficiency of PASS}
Due to its low-loss nature, \acp{pass} are generally considered to have high \ac{ee}. Motivated by this intuitive observation, a few recent studies have investigated the \ac{ee} of \acp{pass}, e.g. \cite{EE_GNN,zeng2025energy}. The lines of work in this respect consider a classical power distribution model for \acp{pass}. This means that they implicitly treat each pinching element as if it is an active antenna whose radiated power can be fully controlled by the transmitter. Considering the \ac{em} radiation in waveguides, one can observe that this is not necessarily the case in \acp{pass} \cite{wang2025modeling}. In fact, the transmit power of each pinching element is explicitly described by its location and coupling characteristics, and hence is less controllable as compared with conventional movable-antenna technologies \cite{mei2024movable}.  

This work aims to develop a realistic formulation for \ac{ee} in \acp{pass}. Motivated by the study in \cite{wang2025modeling}, we formulate the energy consumption and power distribution in a \ac{pass} in terms of the pinching locations and couplings. The power allocation is then addressed implicitly in these systems through location and coupling tuning at the transmitter. 


\subsection{Contributions and Organization}
In this work, we study both \textit{dynamic} and \textit{static} designs for energy efficient transmission in \acp{pass}. Our key contributions are three-fold:
    ($i$) We model the \ac{ee} of \acp{pass} taking into account the \ac{em} propagation in these systems. We show that conventional formulation based on earlier power distribution models is not applicable to \acp{pass}. In fact, the power allocation in these systems is implicitly controlled through tuning of the pinching couplings and locations.
    ($ii$) Using our \ac{ee} model, we formulate energy-efficient transmission in \acp{pass}. The design variables in this problem vary at different rates: while per-user power allocation can be tuned at symbol-level, the coupling coefficients of the elements can be tuned only statically. We hence consider two realistic design scenarios: ($i$) a \textit{dynamic} scenario, in which both \textit{user} power allocation and pinching locations are adjusted at the rate of network coherence, i.e., they change as users change their locations, and the coupling lengths of the elements are tuned statistically. ($ii$) A \textit{static} case in which only the user power allocation is adjusted at the coherence rate, and the pinching locations and couplings are tuned for optimal \textit{ergodic} \ac{ee}.
    ($iii$) We develop a two-tier algorithmic framework, whose inner layer updates the dynamic variables at the coherence rate, while the outer layer tunes the static parameters. Through extensive numerical experiments, we validate our analysis. Our numerical experiments depict a significant throughput gain achieved by dynamic tuning of pinching locations, suggesting high efficiency of \acp{pass} in environment with slow dynamics, e.g. indoor settings with low user mobility.  

\section{System Model}
\label{sec:model}

Consider a dielectric waveguide of length \( L \) with \( N \) pinching elements positioned along the \( x \)-axis at height $a$. Let \( 0 \leq x_{n} \leq L \) be the location of element $n\in \dbc{N}$. The coordinate of the pinching element $n$ can hence be written as $\bp_n = \dbc{x_n, 0, a}$. Due to physical restrictions, each of the two neighboring elements is distanced more than a minimum threshold $\Delta$, i.e., $x_n - x_{n-1} \geq \Delta$. Element $n$ is of length $\ell_n$ and performs as a signal illuminator that operates by coupling with the waveguide. In the sequel, we assume that $x_n$ denotes the \textit{end point} of the element $n$, often called its \textit{effective location}, i.e., the point where signal radiates from. We further denote the location where the element starts coupling with the waveguide by $x_{0,n}$, i.e., $x_n = x_{0,n}+\ell_n$. For sake of simplicity, we consider the following assumptions:
    (1) The propagation loss in the waveguide is negligible.
    (2) The pinching elements and dielectric waveguide have the same refractive indices $i_{\rm ref}$.
    (3) The length of the pinching elements is designed such that complete coupling occurs, i.e., the the coupled power completely transfers to the element and is radiated in open space.
    (4) There is no power reflection at the end of the waveguide, and the non-radiated power at the end of the waveguide is burned to heat. Note that this assumption is realistic, as in practice (as we shall see), with large enough number of pinching elements, this power is negligible.

The waveguide is connected to an \ac{ap}, which deploys this \ac{pass} along with \ac{tdma} for downlink transmission in a multiple access channel with $K$ users. For this setting, we aim to design the system parameters, i.e., the location of elements and their length, such that the \ac{ee} is maximized, i.e., to maximize the average spectral efficiency achieved per unit of power. This requires a concrete characterization of energy consumption in this setting, which we discuss next.

\subsection{Power Model for Downlink PASS}

Using \ac{tdma}, the \ac{ap} transmits $K$ different encoded signals in $K$ time intervals. Let $s_{k,0}$ with power \( P^{\rm in}_k = \abs{s_{k,0}}^2 \) denote the signal fed to the waveguide in time interval $k$ at location $x=0$. The phase-shifted signal that specifies the \ac{em} field at location $x$ of the waveguide in this time interval is given by
\begin{align}
s_{k}\brc{x} = e^{-j \beta x} s_{k,0},    
\end{align}
where $\beta = 2\pi i_{\rm ref} / \lambda$ with $\lambda$ being the wavelength. 

At pinching element $n$, the \ac{em} wave starts coupling at location $x_{0,n}$ and splits into two components by the end of the complete coupling at $x_n$: ($i$) a component that is coupled with the pinching element and radiated in the free space, and ($ii$) a component that is remained in the waveguide. This power splitting repeats towards the last element on the waveguide and specifies the radiated power on each element \cite{wang2025modeling}. To characterize the power radiation in the \ac{pass}, let us start with the first element. Following the analysis in \cite{wang2025modeling}, the signal radiated by the first element is given by
\begin{align}
s_{k,1} = \sin\brc{\kappa \ell_1 } e^{-j \beta x_{1} } s_{k,0},    
\end{align}
for a constant $\kappa$ that represents the coupling coefficient. The signal-carrying components of the remained wave in the waveguide is further given by
\begin{align}
s_{k,1}^{\rm r} = \cos\brc{\kappa \ell_1 } e^{-j \beta x_{1}} s_{k,0} .
\end{align}
The signal $s_{k,1}^{\rm r}$ can be treated as a new input fed at location $x_1$, whose splitting at $x_2-x_1$ specifies the radiation on pinching element $2$. By repeating this procedure, we can conclude that the signal radiated by element $n$ in time interval $k$ is \cite{wang2025modeling}
\begin{align}\label{eq:signal}
    s_{k,n} =
    \sqrt{1-\delta_n^2} \brc{\prod_{i=1}^{n-1} \delta_i}
    e^{-j \beta x_{n}} s_{k,0} ,
\end{align}
where $\delta_i = \cos(\kappa \ell_i)$. In the sequel, we refer to $\delta_i$ as the \textit{power split coefficient} of element $i$. For $\bdelta = \dbc{\delta_1, \ldots, \delta_N}^\trp \in \dbc{0,1}^N$, we further define the effective split coefficient $n$ as
\begin{align}
 A_n\brc{\bdelta} = \sqrt{1-\delta_n^2} \brc{\prod_{i=1}^{n-1} \delta_i},
\end{align}
and write compactly $s_{k,n} = A_n\brc{\bdelta} e^{-j \beta x_{n}} s_{k,0}$, in the sequel.

Considering the \ac{em} propagation, the power radiated from the element $n$ in the time interval $k$ is 
\begin{align}
    P_{k,n}^{\rm tx} =  A_n^2\brc{\bdelta} \abs{s_{k,0}}^2 = 
    \brc{1-\delta_n^2} \brc{\prod_{i=1}^{n-1} \delta_i^2}
    P_k^{\rm in}.
\end{align}
The total transmit power in this interval is thus given by
\begin{align}
    P_{k}^{\rm tx} = \sum_{n=1}^N P_{k,n}^{\rm tx} = P_k^{\rm in} \sum_{n=1}^N 
    \brc{1-\delta_n^2} \brc{\prod_{i=1}^{n-1} \delta_i^2}.
\end{align}

Assuming same time duration for all \ac{tdma} intervals, the average transmitted power in downlink is given by
\begin{align}
    \bar{P}^{\rm tx} = \frac{1}{K} \sum_{k=1}^K P_{k}^{\rm tx} = \bar{P}^{\rm in} \sum_{n=1}^N  A_n^2\brc{\bdelta},
\end{align}
where $\bar{P}^{\rm in} = \sum_{k} P_k^{\rm in} / K$ is the average input power. 

\begin{remark}
As mentioned, we assume that the remaining power in the waveguide is negligible. In asymmetric architectures, one can set this power to zero by matching the last pinching element, i.e., by setting $\delta_N = \cos\brc{\kappa \ell_N} = 0$. In a symmetric architecture, i.e., $\delta_n = \delta$ for all $n\in \dbc{N}$, this is not feasible. However, the remaining power in this case is $\delta^{2N} \bar{P}^{\rm in}$ which can always set arbitrarily close to zero if the length of the elements (specifying $\delta$) is tuned properly. 
\end{remark}

In addition to the input power, there is a circuit power consumed by the radio frequency chain deployed to radiate the modulated signal in the waveguide. This is modeled as a single term $P^{\rm cir}$, as there is a single chain connected to the waveguide. Assuming this term to be fixed, we can model the total power consumed by this \ac{pass} as
\begin{equation}
    P^{\rm total} = \bar{P}^{\rm in} + P^{\rm cir} = \frac{1}{K} \sum_{k=1}^K P_k^{\rm in} + P^{\rm cir}.
\end{equation}
Note that the consumed power in this case does not scale with the \ac{pass} size $N$. This is indeed intuitive: regardless of $N$, the power consumed by the \ac{pass} is specified by the power fed to the waveguide and what is burned in the circuitry. 

\subsection{Characterizing Energy Efficiency}
Considering \ac{em} wave propagation, the signal received by user $k$ at point $\bu_k = [x^{\rm u}_k, y_k^{\rm u},z_k^{\rm u}]$ is given by
\begin{align}
    y_k &= \xi_k
    \sum_{n=1}^{N} \frac{e^{-j \alpha \mathcal{D}_k \brc{x_n}}}{\mathcal{D}_k \brc{x_n}}  s_{k,n} + \varepsilon_k    
\end{align}
where $\varepsilon_k$ is additive Gaussian noise with mean zero and variance $\sigma_k^2$, $\xi_k$ is an attenuation coefficient capturing shadowing, scalar $\alpha = 2\pi/\lambda$ is the wave-number, and $\mathcal{D}_k \brc{x_n}$ denotes the distance between the user and pinching element $n$, i.e., 
\begin{align}
    \mathcal{D}_k \brc{x_n} 
    &= \sqrt{(x_k^{\rm u} - x_{n})^2 + y_k^{\rm u 2} + (z_k^{\rm u}-a)^2}.
\end{align}

Let us define $\bx = \dbc{x_1, \ldots, x_N} \in \dbc{0,L}^N$. Considering \eqref{eq:signal}, we can write compactly $y_k = h_k \brc{\bx, \bdelta} s_{k,0} + \varepsilon_k$ with
\begin{align}
h_k \brc{\bx, \bdelta} = 
    \xi_k
    \sum_{n=1}^{N} \frac{A_n\brc{\bdelta}}{\mathcal{D}_k \brc{x_n}}   e^{-j \brc{\alpha \mathcal{D}_k \brc{x_n} + \beta x_n}},
\end{align}
which describes an effective Gaussian channel from the \ac{ap} to user $k$. We can hence write the downlink transmission rate to user $k$ as
\begin{align}
    \R_k\brc{P_k^{\rm in}, \bx, \bdelta} = \log \brc{1+\Gamma_k P_k^{\rm in} G_k\brc{\bx, \bdelta}},
\end{align}
where $\Gamma_k = \xi_k^2/\sigma_k^2$, and $G_k\brc{\bx, \bdelta}={\abs{h_k \brc{\bx, \bdelta}}^2 }/{\xi_k^2}$ denotes the \textit{effective channel gain} to user $k$. 
Noting that the system operates in \ac{tdma}, the spectral efficiency is given by 
\begin{align}
\se\brc{\mP^{\rm in},\bx, \bdelta} = \frac{1}{K} \sum_{k=1}^K \R_k\brc{P_k^{\rm in}, \bx, \bdelta} .
\end{align}
where $\mP^{\rm in} = \dbc{ P_1^{\rm in}, \ldots, P_K^{\rm in}}^\trp$ is the \textit{user} power allocation. 

Considering the power model and spectral efficiency, the \ac{ee} of the \ac{pass} is given by
\begin{subequations}
\begin{align}
    \mathcal{E}&\brc{\mP^{\rm in}, \bx, \bdelta} 
    = \frac{\se\brc{\mP^{\rm in}, \bx, \bdelta}}{P^{\rm total}} \\
    &= \brc{\sum_{k=1}^K \R_k\brc{P_k^{\rm in}, \bx, \bdelta}}\brc{\sum_{k=1}^K P_k^{\rm in} + K P^{\rm cir}}^{-1} .\label{eq:EE}
\end{align}
\end{subequations}

The expression in \eqref{eq:EE} demonstrates two intuitive facts: ($i$) for a fixed power allocation policy among users, optimizing the \ac{ee} is equivalent to optimization of th spectral efficiency. This is intuitive, as the radiated power in the \ac{pass} is controlled by the position and coupling length of the pinching elements which impact the \ac{ee} only through spectral efficiency. ($ii$) For a fixed \ac{pass} configuration, i.e., fixed $\bx$ and $\bdelta$, the \ac{ee} optimization is equivalent to that of conventional \ac{tdma} systems. This follows the fact that \ac{pass} only provides extra degrees of freedom to modify the communication channel, and does not impact the nature of wave propagation. 

\section{Static and Dynamic Design for PASS}
The ultimate goal of this work is to develop an algorithmic approach to optimize the \ac{ee} of this system. This is a \textit{hybrid} design problem in which the signal-level parameters, i.e., user power allocation, and system-level parameters, i.e., location and coupling length of pinching elements, are designed jointly for optimal \ac{ee}. 
One can readily see that the optimal design depends on the system setting, e.g., users' location. Therefore, any algorithmic approach should update the design parameters at a certain rate. While signal-level parameters can be updated at high rate, the update rate of system-level parameters can be limited due to physical restriction.

In this section, we consider two cases; namely, the \textit{dynamic} and \textit{static} designs. The former refers to the case, where \ac{pass} physical degrees of freedom can be updated at a high rate, while the latter considers the scenarios, in which the system parameters are updated at significantly lower rate.

\subsection{Dynamic Design of PASS}
In dynamic design, we assume that the pinching locations are dynamically tuned at a rate comparable to channel coherence. In other words, the \ac{pass} is able to tune the location of its elements, as the users change their locations. Such design can be realized by implementing each element via multiple mechanically shifting coupling units on the waveguide, each covering one part of the waveguide \cite{ding2024flexible,LiuPASS}. 

Although a dynamic change of pinching locations is realistic in slow changing settings, a dynamic tuning of split coefficients $\delta_n$ is not feasible in practice, as they are controlled by the coupling length of the elements. We thus formulate the energy-efficient design in a dynamic setting as follows.

\begin{definition}[Dynamic EE Design]
The dynamic energy-efficient design of \ac{pass} solves the following problem
\begin{align}\label{Dynam}
    \max_{\bdelta} \;
    &\Ex{
    \max_{\bx , \mP^{\rm in} } 
    \mathcal{E}\brc{\mP^{\rm in}, \bx, \bdelta} 
    }{\bu_1, \ldots, \bu_K} 
    \tag{$\mathcal{D}$}
    \\
    &\subto \bdelta \in \dbc{0,1}^N, \bx \in \setL, \text{ and } \sum_{k=1}^K P_k^{\rm in} \leq K P_0,
    \tag{$\mathrm{C}$} \label{eq:Const}
\end{align}
for some given $P_0$ that specifies the limit on the average power radiated by the \ac{ap} to the waveguide, and $\setL$ that is  defined as
\begin{align}\label{eq:L}
    \mathbb{L} = \set{\bx\in\setR^N: 0\leq x_n \leq L \; \text{and} \; \abs{x_n - x_{n-1}} \geq \Delta},
\end{align}
specifying the set of feasible locations on the waveguide.
\end{definition}

In the dynamic formulation, the user power allocation and pinching locations are optimized for given user locations, while the split coefficients are designed to maximize the \textit{ergodic} \ac{ee}, averaged over user locations. 

\subsection{Static Design of PASS}
In static settings, the update rate of the antenna locations is not in the order of the network coherence. This means that the antenna locations are updated at significantly lower rate as compared to the environmental parameters, e.g. user locations. In this case, the efficient design considers ergodic efficiency of the system. An energy-efficient static design for \ac{pass} is hence formulated as follows. 
\begin{definition}[Static EE Design]
The static energy-efficient design of \ac{pass} solves the following problem
\begin{align}\label{Stat}
    \max_{\bdelta, \bx} \;
    &\Ex{
    \max_{\mP } 
    \mathcal{E}\brc{\mP^{\rm in}, \bx, \bdelta} 
    }{\bu_1, \ldots, \bu_K} 
    \tag{$\mathcal{S}$} 
    \subto \; (\mathrm{C}),
\end{align}
for some average power $P_0$, and $\setL$ given in \eqref{eq:L}. 
\end{definition}

It is worth noting that the key difference between the two designs is the objective with respect to the pinching locations: while a dynamic \ac{pass} updates the locations frequently to optimize the \textit{instantaneous} \ac{ee}, the static \ac{pass} tunes its locations to optimize its \textit{ergodic} throughput. 

\section{Algorithms for Dynamic and Static Design}
We develop a hybrid algorithmic scheme based on block coordinate descent, which approximates the solution by alternating between marginal problems in $\bdelta$, $\mP^{\rm in}$ and $\bx$. Note that in both designs, we have both dynamic and static variables: $\bdelta$ and $\mP^{\rm in}$ are static and dynamic, respectively, while $\bx$ changes from dynamic in \eqref{Dynam} to static in \eqref{Stat}. As a result, our scheme is two-tier: the inner-loop updates the dynamic variables, while static variables are updated in an outer loop that is run at the end of a time window including multiple coherence frames. 

\subsection{Power Allocation}
The user power allocation, i.e. optimization in $\mP^{\rm in}$, is a dynamic program. The marginal problem in this case is specified by treating the other variables as fixed: let
    $a_k = \Gamma_k G_k\brc{\bx, \bdelta}$.
Then, the power allocation task reduces to
\begin{align}
    \max_{\mP^{\rm in}} \frac{\sum_k \log \brc{1+a_k P_k^{\rm in}}}{K P^{\rm cir} + \sum_k P_k^{\rm in}} \subto \sum_k P_k^{\rm in} \leq P_0.
    \tag{$\mathcal{M}_1$}
    \label{eq:M1}
\end{align}
The marginal problem in \eqref{eq:M1} describes a standard \ac{ee} maximization, whose solution is given by Dinkelbach method \cite{EE_WF}. Starting from an initial $\lambda\in\setR$, we iterate as follows: 
\begin{enumerate}
    \item For a fixed $\lambda$, solve
    \begin{align}
    &\max_{\mP^{\rm in}} \dbc{{\sum_k \log \brc{1+a_k P_k^{\rm in}}}-\lambda 
    P_k^{\rm in}} \subto \sum_k P_k^{\rm in} \leq P_0. \nonumber
\end{align}
\item Use the solution of Step 1 to update $\lambda$ as
\begin{align}
    \lambda \leftarrow \frac{\sum_k \log \brc{1+a_k P_k^{\rm in}}}{K P^{\rm cir} + \sum_k P_k^{\rm in}}.
    \nonumber
\end{align}
\end{enumerate}
The solution to the optimization in Step 1 is given by water-filling. The power allocation algorithm describes an iterative water-filling, as summarized in Algorithm~\ref{alg1}.
\begin{algorithm}[t]
\caption{User Power Allocation}
\label{alg1}
\begin{algorithmic}[1]
\State Initialize $\lambda \gets 0$
\Repeat
    \Statex \texttt{\#apply water-filling}
    \State Find $\ell \geq \lambda$ by checking the slackness
    \State Update $P_k^{\rm in} \gets \max\{0,\,1/\ell - 1/a_k\}$
    \State Update $\R \gets \sum_k \log(1+a_k P^{\rm in}_k)$ and $\mathcal{P} \gets \sum_k P^{\rm in}_k$
    \Statex \texttt{\#update $\lambda$}
    \State $\phi \gets \R - \lambda \brc{K P^{\rm cir}+\mathcal{P}}$
    \State $\lambda \gets \R/ \brc{K P^{\rm cir}+\mathcal{P}}$
\Until{$\abs{\phi} \le \varepsilon$}
\end{algorithmic}
\end{algorithm}

\subsection{PASS Tuning}
The \ac{pass} tuning problem, i.e. optimization in $\bx$, changes from dynamic in \eqref{Dynam} to static in \eqref{Stat}. We hence discuss the marginal problem of each design separately in the sequel.  
\paragraph*{Dynamic Case} In dynamic design, for fixed $\mP^{\rm in}$ and $\bdelta$, we marginally solve
\begin{align}
     \max_{\bx\in \setL} \sum_k 
    \log \brc{1+\Gamma_k P_k^{\rm in} G_k\brc{\bx, \bdelta}}.
        \tag{$\mathcal{M}^{\cal D}_2$}
    \label{eq:DM2}
\end{align}
Due to the extreme fluctuation on the objective surface, the classical gradient decent leads to an unreliable solution to \eqref{eq:DM2}. We therefore solve this problem via Gauss-Seidel approach: let $x_r = x$ denote the location of element $r$. Starting with $r=1$, we update $x_r$, while treating the other elements as fixed. In this case, $G_k\brc{\bx, \bdelta}$ is given in terms of $x_r$ as
\begin{align}
\hspace{-2mm}G_k\brc{\bx, \bdelta} = f_{k,r} \brc{x} = \abs{ 
    \tau_{k,r}
    +
    \frac{A_r\brc{\bdelta}}{\mathcal{D}_k \brc{x}}   e^{-j \brc{\alpha \mathcal{D}_k \brc{x} + \beta x}} 
    }^2 \hspace{-2mm},
\end{align}
where $\tau_{k,r}$ is given by
\begin{align}
      \tau_{k,r }=  \sum_{n\neq r}  
    \frac{A_n\brc{\bdelta}}{\mathcal{D}_k \brc{x_n}}   e^{-j \brc{\alpha \mathcal{D}_k \brc{x_n} + \beta x_n}}.
\end{align}
Replacing in \eqref{eq:DM2}, the tuning of element $r$ reduces~to~optimizing the scalar objective 
\begin{align}
     F_r \brc{x} = \sum_k 
    \log \brc{1+\Gamma_k P_k^{\rm in} f_{k,r} \brc{x}},
\end{align}
which can be addressed by search on a fine grid. The final algorithm is summarized in Algorithm~\ref{alg2}.
\begin{algorithm}[t]
\caption{Pinching Location Tuning}
\label{alg2}
\begin{algorithmic}[1]
\State Initialize $\bx \in \setL$ and a grid set $\mathbb{G}$ on $\dbc{0,L}$ 
\For{$r=1:N$}
\State Find $f_{k,r} \brc{x}$ and compute $\mathbb{F} = \set{F_r \brc{x}: x\in \mathbb{G}}$
\State Update $x_r \gets \argmax \mathbb{F}$
\While{there is $n < r$ such that $\abs{x_r - x_n} <  \Delta$}
    \State Update $\mathbb{F} = \mathbb{F} - \set{F_r \brc{x_r}}$ and $x_r \gets \argmax \mathbb{F}$
\EndWhile
\EndFor
\State $\bx \gets \texttt{sort}\dbc{x_1, \ldots,x_N}$
\end{algorithmic}
\end{algorithm}

\paragraph*{Static Case} The \ac{pass} tuning in the static design reduces to the following marginal optimization
\begin{align}
     \max_{\bx\in \setL} \Ex{
     \frac{
     \sum_k 
    \log \brc{1+\Gamma_k P_k^{\rm in} G_k\brc{\bx, \bdelta}}
    }{
    K P^{\rm cir} + \sum_k P_k^{\rm in}
    }
    }{\bu_1, \ldots, \bu_K}.
            \tag{$\mathcal{M}^{\cal S}_2$}
    \label{eq:SM2}
\end{align}
It is worth noting that unlike the dynamic case, this optimization is not generally equivalent to SE maximization, since the power allocation is dynamically changing. 

Assuming that the coherence interval of the system, i.e. number of transmission frames, per which the location of the elements is optimized once, contains $m$ samples of locations $\bu_{k,1}, \ldots, \bu_{k,m}$ for user $k$, we can estimate the ergodic \ac{ee} as
\begin{align}
     \hat{\mathcal{E}}\brc{\bx} = \frac{1}{m} \sum_{i=1}^m 
     \frac{\sum_k \log \brc{1+\Gamma_{k} P_{k,i}^{\rm in} G_{k,i}\brc{\bx, \bdelta} }}{
    K P^{\rm cir} + \sum_k P_{k,i}^{\rm in}
    }  ,
\end{align}
where $G_{k,i}\brc{\bx, \bdelta}$ and $P_{k,i}^{\rm in}$ denote the channel gain in sample location $\bu_{k,i}$ and its corresponding dynamic power allocation. 

Using the estimate $\hat{\mathcal{E}}\brc{\bx}$, \eqref{eq:SM2} is approximately solved by
\begin{align}
     \max_{\bx\in \setL} \hat{\mathcal{E}}\brc{\bx},
            \tag{$\hat{\mathcal{M}}^{\cal S}_2$}
    \label{eq:hatSM2}
\end{align}
which is solved via the Gauss-Seidel approach: defining 
\begin{align}
f_{k,r,i} \brc{x} = \abs{ 
    \tau_{k,r,i}
    +
    \frac{A_r\brc{\bdelta}}{\mathcal{D}_{k,i} \brc{x}}   e^{-j \brc{\alpha \mathcal{D}_{k,i} \brc{x} + \beta x}} 
    }^2,
\end{align}
with $\tau_{k,r,i}$ being
\begin{align}
      \tau_{k,r,i}=  \sum_{n\neq r}  
    \frac{A_n\brc{\bdelta}}{\mathcal{D}_{k,i} \brc{x_n}}   e^{-j \brc{\alpha \mathcal{D}_{k,i} \brc{x_n} + \beta x_n}},
\end{align}
we compute the scalar objective for element $r$ as 
\begin{align}
     F_r \brc{x} = \frac{1}{m} \sum_{i=1}^m 
     \frac{1}{P^{\rm sum}_i} {\sum_k \log \brc{1+\Gamma_{k} P_{k,i}^{\rm in} f_{k,r,i} \brc{x} }},
\end{align}
where $P^{\rm sum}_i$ is defined as
\begin{align}
    P^{\rm sum}_i = K P^{\rm cir} + \sum_k P_{k,i}^{\rm in}.
\end{align}
We solve each scalar problem by a linear search on a fine grid.

\subsection{Tuning Coupling Length}
The coupling tuning is performed statically by solving 
\begin{align}
     \max_{\bdelta } \Ex{
     \frac{
     \sum_k 
    \log \brc{1+\Gamma_k P_{k}^{\rm in} G_k\brc{\bx, \bdelta}}
    }{
    K P^{\rm cir} + \sum_k P_k^{\rm in}
    }
    }{\bu_1, \ldots, \bu_K}.
               \tag{${\mathcal{M}}_3$}
    \label{eq:M3}
\end{align}
The objective in this case is a well-defined function in $\bdelta$, and hence we can address this task by gradient ascent. 

Similar to the static design of \ac{pass}, we assume that we have $m$ samples of user locations in a coherence interval. Let the user allocation vector and antenna locations at sample $i$ be $\mP^{\rm in}_i$ and $\bx_i$, respectively. Note that in the dynamic location tuning $\bx_i \neq \bx_j$ in general for two samples $i\neq j$, while in the static case, $\bx_1 = \ldots= \bx_m$. The objective is then estimated~as 
\begin{align}
     \hat{\mathcal{E}} \brc{\bdelta} = \frac{1}{m} \sum_{i=1}^m 
     \frac{1}{P^{\rm sum}_i} {\sum_k \log \brc{1+\Gamma_{k} P_{k,i}^{\rm in} G_{k,i}\brc{\bx_i, \bdelta} }}.
\end{align}
This defines a smooth function in $\bdelta$, whose maximum is tracked via gradient ascent. To this end, we write 
\begin{align}
      \nabla \hat{\mathcal{E}}  = \frac{1}{m} \sum_{i=1}^m 
     \frac{1}{P^{\rm sum}_i} {\sum_k \frac{\Gamma_{k} P_{k,i}^{\rm in} \nabla G_{k,i}\brc{\bx_i, \bdelta} }{1+\Gamma_{k} P_{k,i}^{\rm in} G_{k,i}\brc{\bx_i, \bdelta} }}.
\end{align}
Using chain rule,  
       $\nabla G_{k,i}\brc{\bx_i, \bdelta} = 2 \Re\set{ \Xi_{k,i}^* \brc{\bdelta} \nabla \Xi_{k,i} \brc{\bdelta} }$, where we define
\begin{align}
   \Xi_{k,i} \brc{\bdelta} = \sum_r \eta_{k,r,i} A_r \brc{\bdelta}
\end{align}
with $\eta_{k,n,i}$ being
\begin{align}
    \eta_{k,n,i} =   
    \frac{
    e^{-j \brc{\alpha \mathcal{D}_{k,i} \brc{x_{i,n}} + \beta x_{i,n}}}
    }{\mathcal{D}_{k,i} \brc{x_{i,n}}}.
\end{align}
The gradient of $\Xi_{k,i} \brc{\bdelta}$ is further computed in terms of the gradient of $A_n\brc{\bdelta}$ using the identity
\begin{align}
       \frac{\nabla A_n\brc{\bdelta}}{A_n\brc{\bdelta}} = \dbc{
       {\frac{1}{\delta_1}, \ldots, \frac{1}{\delta_{n-1}}}, -\frac{\delta_n}{{1-\delta_n^2}}, 0, \ldots, 0
       }^\trp.
\end{align}

Using the above derivation, the local maximizer is determined by tracking the gradient. Nevertheless, using the vanilla gradient ascent, the coupling coefficients can be update to invalid values, i.e., $\delta_n \notin [0,1]$. To this end, we further project the update $\bdelta$ to the feasible region after each update. This concludes the iterative algorithm in Algorithm~\ref{alg3}. 
\begin{algorithm}[t]
\caption{Coupling Tuning}
\label{alg3}
\begin{algorithmic}[1]
\State Initialize $\bdelta \in \dbc{0,1}^N$ and choose a small step size $\mu$
\Repeat
\State Compute $ G_{k,i}\brc{\bx_i, \bdelta}$ and $\nabla G_{k,i}\brc{\bx_i, \bdelta}$ at $\bdelta$
\State Update $\bdelta \gets \bdelta + \mu \nabla \hat{\mathcal{E}}$
\For{$n=1:N$}
\If{$\delta_n > 1$} 
Set $\delta_n = 1$
\Else
\textbf{ if} $\delta_n < 0$ \textbf{then} Set $\delta_n = 0$
\EndIf
\EndFor
\Until{it converges}
\end{algorithmic}
\end{algorithm}

\section{Numerical Experiment}
We next validate the proposed designs through numerical experiments and compare it against the baseline. Unless stated otherwise, $K=6$ users are randomly and uniformly distributed within a rectangular service region of size $D_x=50$ m and $D_y=20$ m, centered at $[{D_x}/{2}, 0, 0]$. The waveguide is aligned along the $x$-axis, deployed at a height of $a = 3$ m, and covers the entire service area. The parameters of the simulation setup are summarized as follows: carrier frequency is $f_{\rm{c}}=28$ GHz, effective refractive index is $i_{\rm ref}=1.4$, circuit power is set to $P^{\rm{cir}}=0$ dBm, grid search resolution is $Q=10^4$, minimum inter-antenna distance is set to $\Delta={\lambda}/{2}$, and noise variance is $\sigma_{k}^2=-90$ dBm. In iterative algorithms, user powers are initialized as $P_k^{\rm{in}}={P}/{K}$ for a total power $P$, the splitting coefficients are set to $\delta_n=0.5$ for $n\in[N]$, and the initial pinching locations are uniformly distributed, i.e. $x_n=(n-1){D_x}/{N-1}$. For comparison, we evaluate the proposed coupling-length–tunable \ac{pass} against a baseline configuration where all antennas employ a fixed power-splitting coefficient of $\delta_n=0.5$ for $n\in[N]$.

\begin{figure}[!t]
\centering
\includegraphics[width=0.4\textwidth]{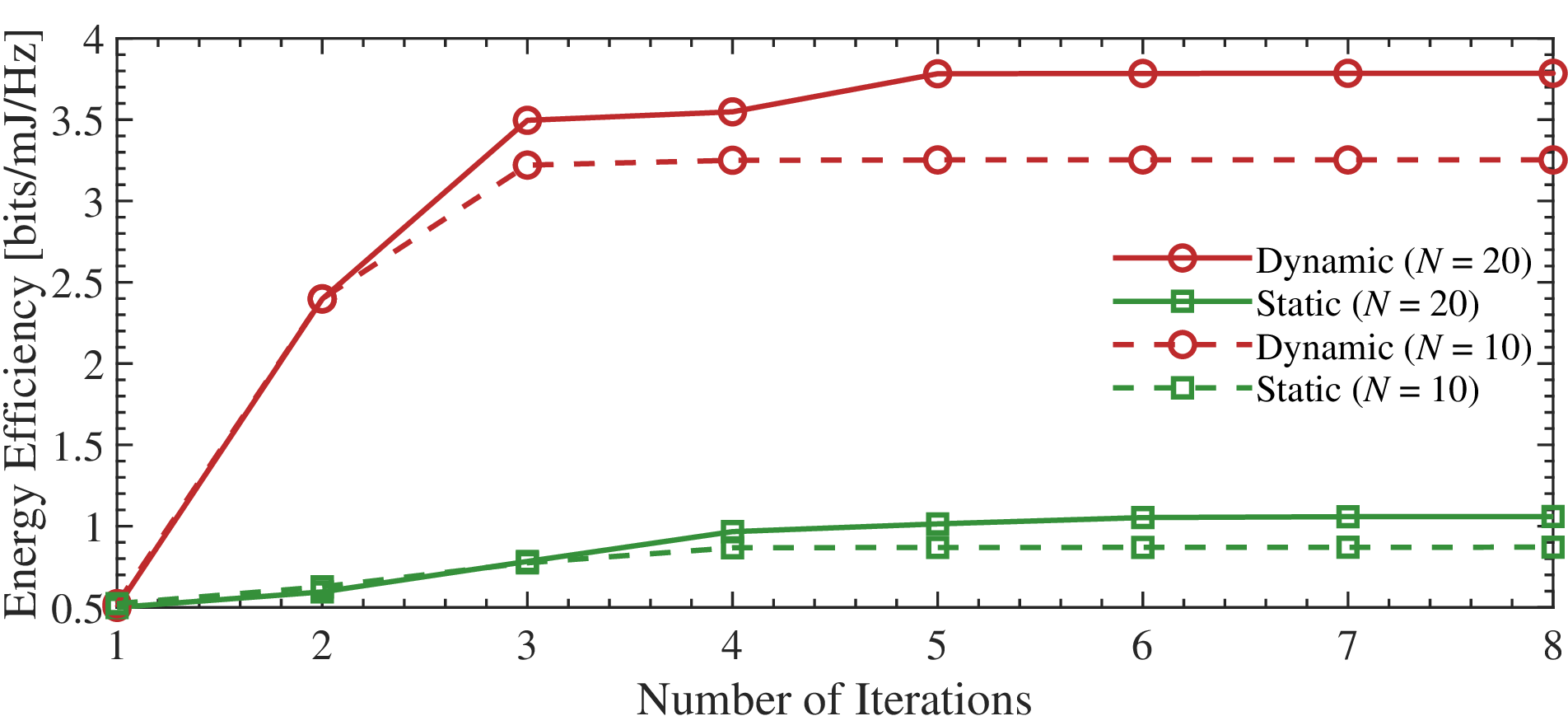}
\caption{Convergence of the proposed algorithms.}
\label{Figure_EE_Convergence}
\vspace{-10pt}
\end{figure}

{\figurename} {\ref{Figure_EE_Convergence}} illustrates the convergence behavior of the proposed algorithmic scheme for both dynamic and static designs under different numbers of antenna elements $N$. As observed, the EE increases with the number of iterations and converges to a stable value, confirming the convergence and effectiveness of the proposed methods. Moreover, the dynamic design achieves considerably higher EE than the static design. This observation is intuitive, as the dynamic approach adaptively optimizes the antenna placement within each coherence interval.

\begin{figure}[!t]
\centering
\includegraphics[width=0.4\textwidth]{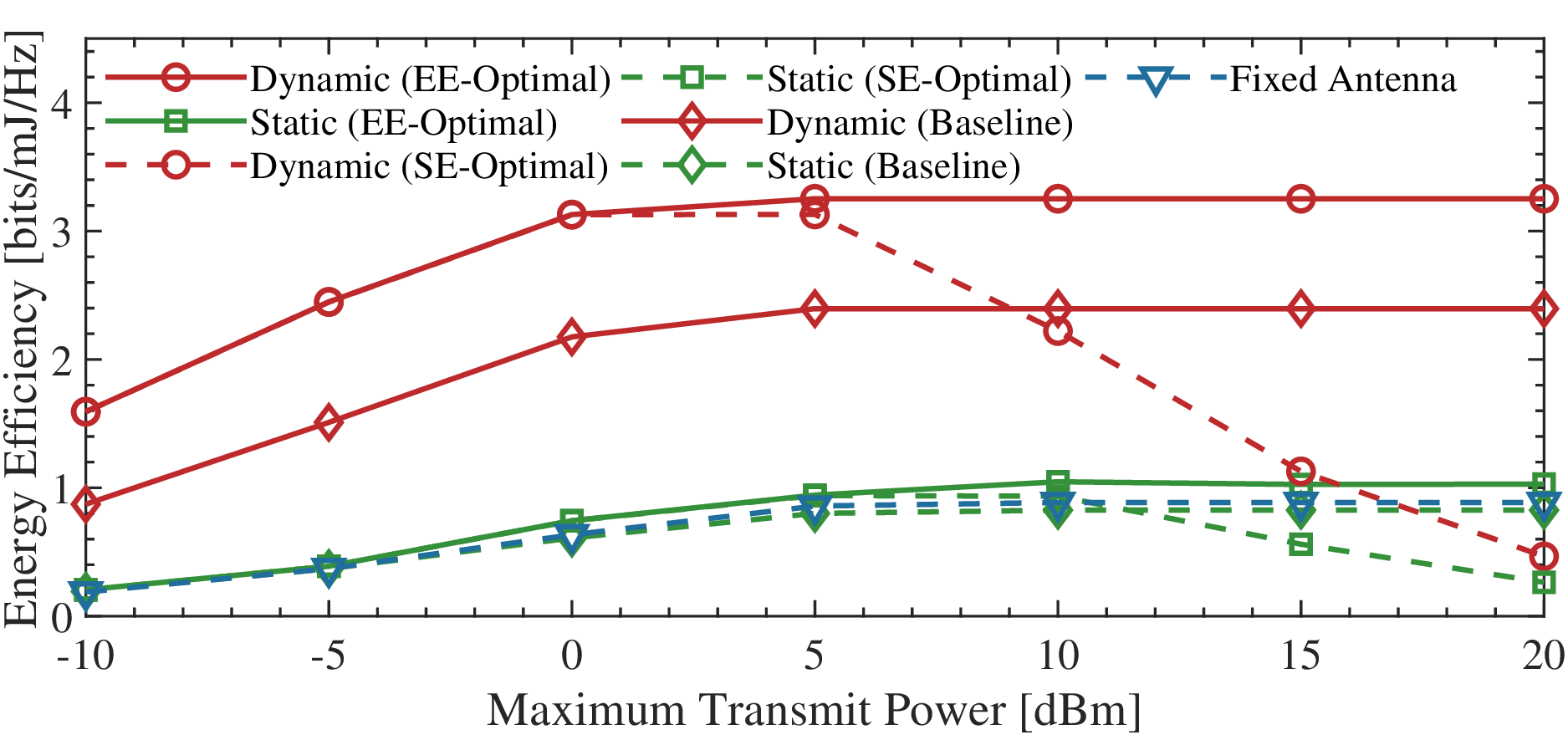}
\caption{EE vs the transmit power at $N=10$.}
\label{Figure_EE_Power}
\vspace{-10pt}
\end{figure}

{\figurename} {\ref{Figure_EE_Power}} shows the EE against the maximum transmit power $P$. For completeness, both EE-oriented and SE-oriented designs are evaluated, where the latter uses the proposed framework to directly maximize the SE. As shown, for both dynamic and static configurations, the EE-oriented design achieves higher EE than the SE-oriented counterpart, particularly in the high-SNR regime. Furthermore, the dynamic design outperforms the static one due to its ability to adaptively optimize the pinching-antenna placement for each user location. This highlights the critical role of antenna placement in enhancing system efficiency. In addition, when compared with the baseline scheme employing a fixed coupling length, the proposed PASS design with tunable coupling length significantly improves the EE especially under dynamic design, which validates the effectiveness of adaptive coupling-length control in improving system performance. For reference, the results for a conventional fixed-position antenna located at the center of the service region are also included. Although the static design avoids frequent updates of antenna positions, its performance gain over traditional fixed-antenna systems remains limited, underscoring the importance of spatial adaptability in PASS.

\begin{figure}[!t]
\centering
\includegraphics[width=0.4\textwidth]{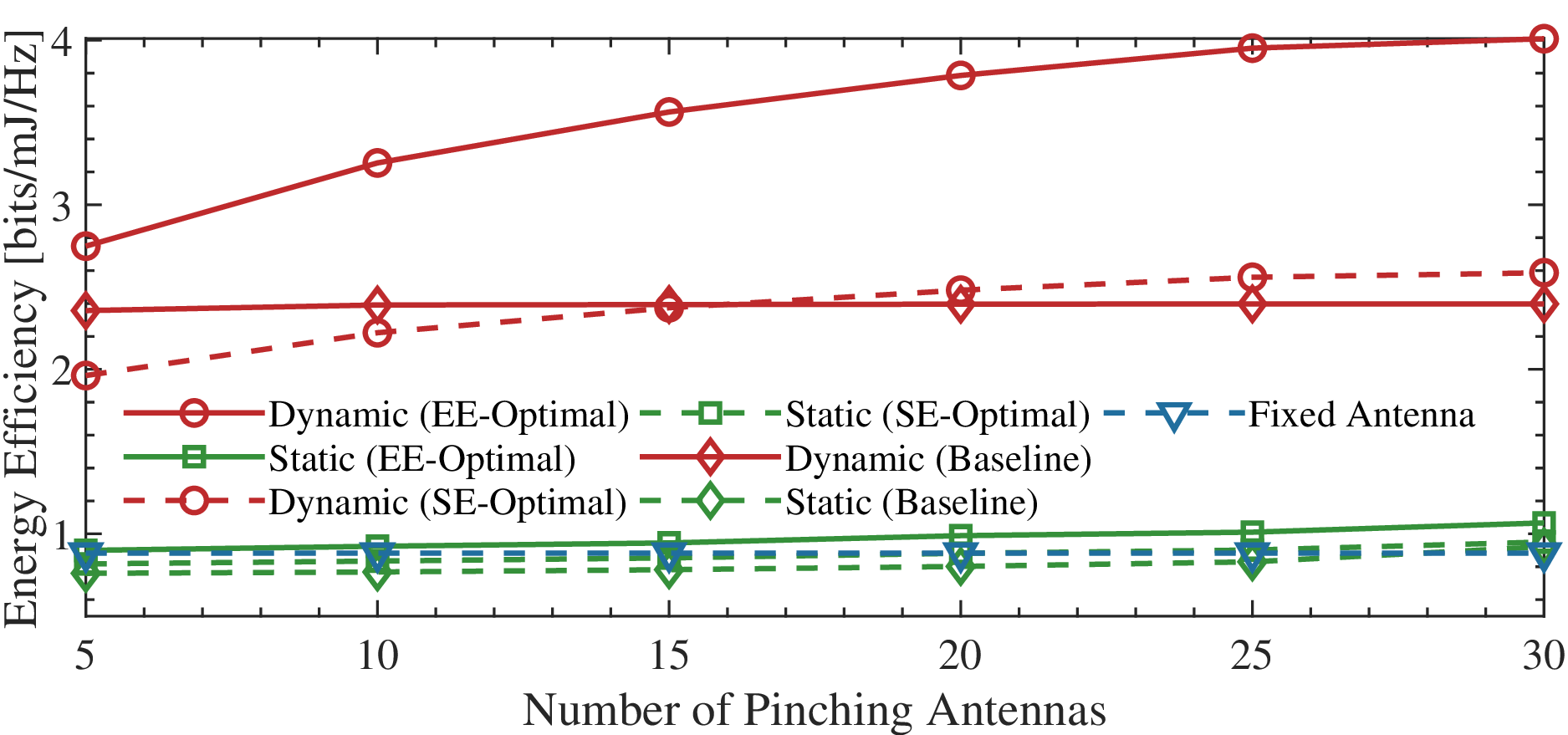}
\caption{EE vs the number of elements at $P_0=10$ dBm.}
\label{Figure_EE_Antenna_Number}
\vspace{-10pt}
\end{figure}

\begin{figure}[!t]
\centering
\includegraphics[width=0.4\textwidth]{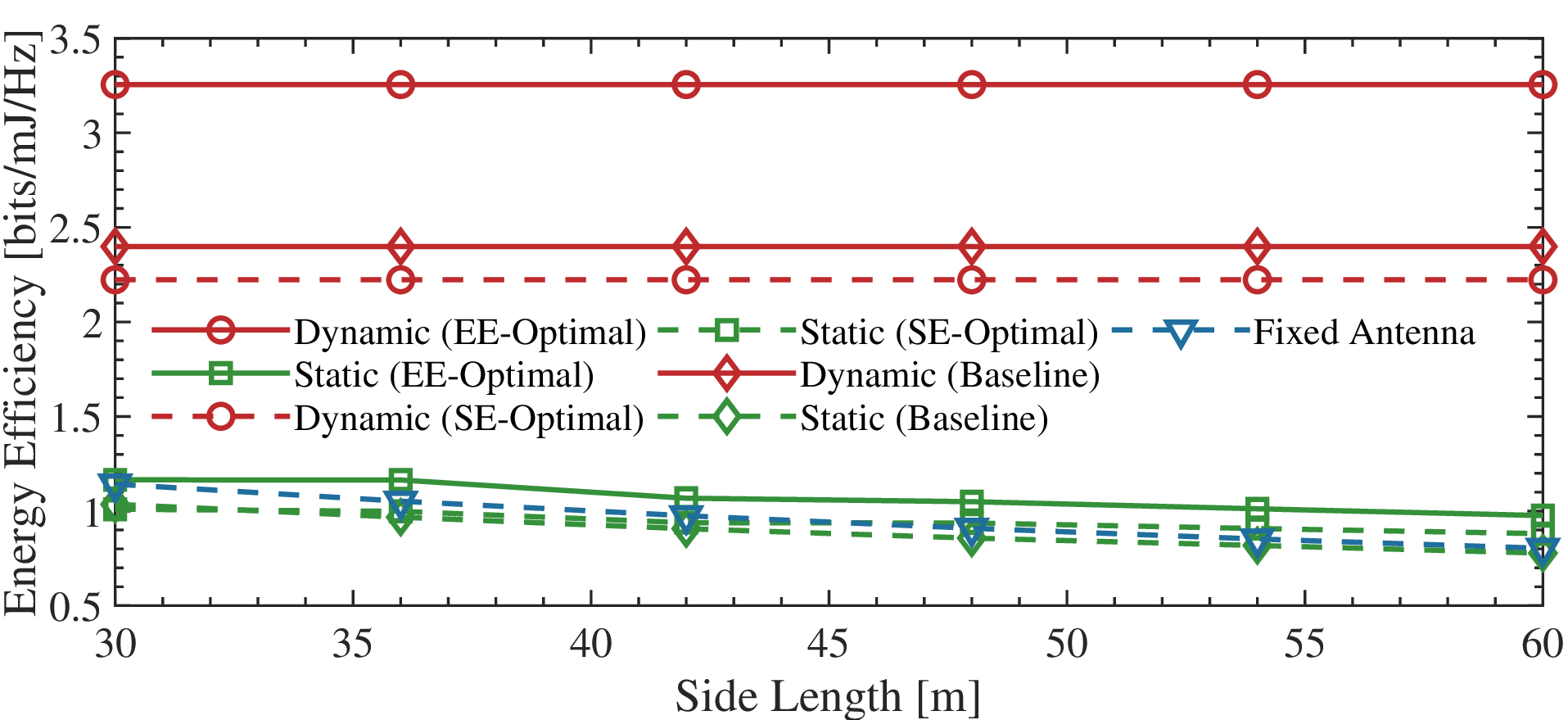}
\caption{EE vs the side length $D_x$ at $P_0=10$ dBm and $N=10$.}
\label{Figure_EE_Side_Length}
\vspace{-10pt}
\end{figure}

{\figurename} {\ref{Figure_EE_Antenna_Number}} plots the EE as a function of the number of antennas. The figure shows that for the proposed designs, the achievable EE increases steadily with the antenna number. In contrast, for the PASS design with tunable coupling length, the EE exhibits only marginal improvement as the number of antennas grows. This occurs because, under fixed coupling length, most of the radiated energy is emitted by the first few antennas, and thus adding more antennas contributes little to overall system efficiency. Finally, in {\figurename} {\ref{Figure_EE_Side_Length}}, we plot the EE against the side length $D_x$. For the dynamic design, the antenna placement is optimized individually for each user within its allocated time slot. This allows the system to maintain nearly constant EE even when users are distributed over a wider area. In contrast, the static design optimizes antenna placement based on the overall user distribution, and its achievable EE decreases as $D_x$ increases. This degradation arises, since the average user-to-antenna distance becomes larger with a wider service region, which leads to higher path loss. 

\section{Conclusions}
This work formulated the \ac{ee} of \acp{pass} considering their \ac{em} characteristics. Unlike conventional movable-antenna technologies, the per-element power distribution in \acp{pass} is not controlled explicitly through an allocation policy, but only implicitly through the pinching locations and couplings. We developed static and dynamic algorithms for \ac{ee} maximization in \acp{pass}. While the proposed designs outperform the baseline in both cases, dynamic tuning of pinching elements enables the \ac{pass} to significantly boost the throughput, as compared with the baseline. This suggests that \acp{pass} can operate significantly energy-efficient in scenarios with slow fluctuations in the network, e.g. indoor environments with low user mobility. Extending the proposed \ac{ee} formulation to \ac{mimo} settings is a natural direction for future work, which is currently ongoing.\vspace{5mm}

\bibliographystyle{IEEEtran}
\bibliography{strings,refs}

@misc{wang2025modeling,
  title={Modeling and beamforming optimization for pinching-antenna systems},
  author={Wang, Zhaolin and Ouyang, Chongjun and Mu, Xidong and Liu, Yuanwei and Ding, Zhiguo},
  Note={arXiv preprint arXiv:2502.05917},
  year={2025}
}

@ARTICLE{ding2024flexible,
  author={Ding, Zhiguo and Schober, Robert and Vincent Poor, H.},
  journal={IEEE Trans. Commun.}, 
  title={Flexible-Antenna Systems: A Pinching-Antenna Perspective}, 
  year={2025},
  volume={},
  number={},
  pages={1-1},
  doi={10.1109/TCOMM.2025.3555866}}

@misc{bereyhi2025mimopass,
      title={{MIMO-PASS}: Uplink and Downlink Transmission via {MIMO} Pinching-Antenna Systems}, 
      author={Ali Bereyhi and Chongjun Ouyang and Saba Asaad and Zhiguo Ding and H. Vincent Poor},
      year={2025},
      note={arXiv preprint arXiv:2503.03117},
}

@ARTICLE{LiuPASS,
  author={Liu, Yuanwei and Wang, Zhaolin and Mu, Xidong and Ouyang, Chongjun and Xu, Xiaoxia and Ding, Zhiguo},
  journal={IEEE Communications Magazine}, 
  title={Pinching-Antenna Systems: Architecture Designs, Opportunities, and Outlook}, 
  year={2025},
  volume={},
  number={},
  pages={1-7},
  doi={10.1109/MCOM.001.2500037}}

@misc{zeng2025energy,
  title={Energy-Efficient Design for Downlink Pinching-Antenna Systems with {QoS} Guarantee},
  author={Zeng, Ming and Wang, Ji and Zhou, Gui and Fang, Fang and Wang, Xianbin},
  note={arXiv preprint arXiv:2505.14904},
  year={2025}
}

@INPROCEEDINGS{EE_WF,
  author={Prabhu, R. S. and Daneshrad, B.},
  booktitle={Proc. IEEE Int. Conf. Commun. (ICC)}, 
  title={An Energy-Efficient Water-Filling Algorithm for {OFDM} Systems}, 
  year={2010},
  volume={},
  number={},
  pages={1-5},
  doi={10.1109/ICC.2010.5502818}}

@ARTICLE{EE_GNN,
  author={Xie, Xinke and Lu, Yang and Ding, Zhiguo},
  journal={IEEE Wireless Commun. Lett.}, 
  title={Graph Neural Network Enabled Pinching Antennas}, 
  year={2025},
  volume={14},
  number={9},
  pages={2982-2986},
  doi={10.1109/LWC.2025.3584919}}

@misc{shan2025securemulti,
      title={Secure Multicast Communications with Pinching-Antenna Systems {(PASS)}}, 
      author={Shan Shan and Chongjun Ouyang and Yong Li and Yuanwei Liu},
      year={2025},
      note={arXiv preprint arXiv:2509.16045},
}

@ARTICLE{PassOut,
  author={Tyrovolas, Dimitrios and Tegos, Sotiris A. and Diamantoulakis, Panagiotis D. and Ioannidis, Sotiris and Liaskos, Christos K. and Karagiannidis, George K.},
  journal={IEEE Trans. Cogn. Commun. Net.}, 
  title={Performance Analysis of Pinching-Antenna Systems}, 
  year={2025},
  volume={},
  number={},
  pages={1-1},
  doi={10.1109/TCCN.2025.3564470}}

@article{suzuki2022pinching,
  title={Pinching Antenna: Using a Dielectric Waveguide as an Antenna},
  author={Fukuda, Atsushi and Yamamoto, Hiroto and Okazaki, Hiroshi and Suzuki, Yasunori and Kawai, Kunihiro},
  journal={NTT DOCOMO Technical J.},
  volume={23},
  number={3},
  pages={5--12},
  year={2022},
}

@misc{ouyang2025array,
  title={Array Gain for Pinching-Antenna Systems ({PASS})},
  author={Ouyang, Chongjun and Wang, Zhaolin and Liu, Yuanwei and Ding, Zhiguo},
  note={arXiv preprint arXiv:2501.05657},
  year={2025}
}

@article{mei2024movable,
  title={Movable-antenna position optimization: A graph-based approach},
  author={Mei, Weidong and Wei, Xin and Ning, Boyu and Chen, Zhi and Zhang, Rui},
  journal={IEEE Wireless Commun. Lett.},
  volume={13},
  number={7},
  pages={1853--1857},
  year={2024},
  month=jul,
  publisher={IEEE}
}

@inproceedings{bereyhi2025icc,
  title={Downlink beamforming with pinching-antenna assisted {MIMO} systems},
  author={Bereyhi, Ali and Ouyang, Chongjun and Asaad, Saba and Ding, Zhiguo and Poor, H. Vincent},
  booktitle={Proc. IEEE Int. Conf. Commun. Workshop (ICC Workshop) },
  pages={1--7},
  year={2025},
}
\end{document}